\documentclass[]{aa}
\usepackage{graphicx} 
\usepackage{txfonts}

\newcounter{subfig}
\newcommand{\skipthis}[1]{}
\newcommand{\hii}{{\rm H}{\sc ii}}

\def\nh3{$\rm{NH_3}$}
\def\NH3{$\rm{NH_3}$}

\def\msun{M$_\odot$}
\def\lsolar{L$_\odot$}
\def\lsun{L$_\odot$}
\def\kms-1{km~s$^{-1}$}
\def\h2o{$\rm{H_2O}$}
\def\h2{$\rm{H_2}$}
\def\cm3{$\rm{cm^{-3}}$}
\def\hcop{$\rm{HCO^+}$}
\def\vlsr{$\rm{V_{LSR}}$}

\newcommand{\lsim}{${\raisebox{-.9ex}{$\stackrel{\textstyle<}{\sim}$}}$ }
\newcommand{\gsim}{${\raisebox{-.9ex}{$\stackrel{\textstyle>}{\sim}$}}$ }

\begin{document}

\title{A Jet-like Outflow toward the High-Mass (Proto)stellar Object IRAS 18566+0408}

\subtitle{}

\author{Qizhou Zhang\inst{1}, T. K. Sridharan\inst{1}, Todd R. Hunter\inst{2},  
Yuan Chen\inst{1}, Henrik Beuther\inst{3} and Friedrich Wyrowski\inst{4}\fnmsep}

\institute{Harvard-Smithsonian Center for Astrophysics,
60 Garden Street,
Cambridge, Massachusetts 02138, USA\\
              \email{qzhang@cfa.harvard.edu} 
\and
National Radio Astronomical Observatory, 
 520 Edgemont Road,
Charlottesville, VA 22903-2475
\and
Max-Planck-Institute for Astronomy,
K\"onigstuhl 17, 69117 Heidelberg, Germany 
\and
Max-Planck-Institut f{\"u}r Radioastronomie,
Auf dem Hugel 69,
53121 Bonn, Germany\\}

\date{Received ...; accepted ...}

\abstract
{Studies of high-mass protostellar objects reveal important
information regarding the formation process of massive stars.}  
{ We study the physical conditions in the dense
core and molecular outflow associated with the high-mass protostellar
candidate IRAS 18566+0408 at high angular resolution.}  
{We performed interferometric observations in the
\nh3 (J,K)=(1,1), (2,2) and (3,3) inversion transitions, the SiO J=2-1
and HCN J=1-0 lines, and the 43 and 87 GHz continuum emission using
the VLA and OVRO. } 
{The 87 GHz
continuum emission reveals two continuum peaks MM-1 and MM-2 along a
molecular ridge. The dominant peak MM-1 coincides with a compact
emission feature at 43 GHz, and arises mostly from the dust
emission. For dust emissivity index $\beta$ of 1.3,
the masses in the dust peaks amount to 70 \msun\ for MM-1, and
27 \msun\ for MM-2. Assuming internal heating,
the central luminosities of MM-1 and MM-2 are $6 \times 10^4$ and $8
\times 10^3$ \lsun, respectively.

The SiO emission reveals a well collimated outflow emanating from
MM-1. The jet-like outflow is also detected in \nh3 at velocities
similar to the SiO emission.  The outflow, with a mass of 27 \msun,
causes significant heating in the gas to temperatures of 70 K, much
higher than the temperature of $\lsim 15$ K in the extended core.
Compact ($< 3''$) and narrow line ($<1.5$ \kms-1) \nh3 (3,3) emission
features are found associated with the outflow.  They likely arise
from weak population inversion in \nh3 similar to the maser emission.

Toward MM-1, there is a compact \nh3 structure with a linewidth
that increases from 5.5 \kms-1\ FHWM measured at 3$''$ resolution to
8.7 \kms-1\ measured at 1$''$ resolution. This linewidth is much
larger than the FWHM of $<$ 2 \kms-1\ in the entire core, and does not
appear to originate from the outflow.  This large linewidth may arise
from rotation/infall, or relative motions of unresolved protostellar
cores.}  
{}


\authorrunning{Zhang et al.} 
\titlerunning{Jet-like Outflow toward IRAS 18566+0408} 
 
\maketitle
 
\section{Introduction}

Systematic surveys in the past decade identified hundreds of high-mass protostellar
candidates (Molinari et al. 1996; Sridharan 2002; Fontani et al. 2005).
These objects, selected initially from the IRAS point source catalog,
typically have far infrared luminosities of $\gsim 10^3$ \lsun, contain
$10^2 - 10 ^4$ \msun\ of dense molecular gas
(Molinari et al. 2002; Beuther et al. 2002a; Williams,
Fuller \& Sridharan 2004;
Beltr{\'a}n et al. 2006), and are associated
with massive molecular outflows (Zhang et al. 2001, 2005; Beuther et al. 2002b).
Compared with ultra compact \hii\ (UC\hii) regions,
high-mass protostellar candidates have similar amounts of dense
molecular gas, but are less luminous and have much weaker emission 
at centimeter wavelengths.
Therefore, they are likely to be in an earlier evolutionary stage than the UC\hii\
phase.

These surveys were carried out mostly using single dish telescopes
with angular resolutions of $> 10''$. High angular resolution imaging
is required to probe dense cores and molecular outflows at spatial
scales relevant to massive protostars.  In the past few years, images
from (sub)mm interferometers have often resolved poorly-collimated
outflows identified by single dish telescopes into multiple
well-collimated outflows (e.g. IRAS~05358+3543, Beuther et al. 2002d;
AFGL~5142, Zhang et al. 2007). In the meantime, high resolution images
in \nh3 and other dense molecular gas tracers reveal interesting
kinematics close to massive protostars (Zhang et al. 1998, 2002).

In this paper, we present a high resolution study toward the high-mass
protostellar candidate IRAS 18566+0408. At a kinematic distance of 6.7
kpc (Sridharan et al. 2002), the source has a far infrared luminosity
of several $ 10^4$ \lsolar.  The object was initially undetected at 2
and 6 cm at an rms of 0.16 mJy and 0.1 mJy, respectively (Miralles,
Rodr{\'\i}guez \& Scalise 1994), but later detected at 3.6 cm at a
flux density of 0.7 mJy (Carral et al. 1999), and at 2cm at a flux of
0.7 mJy (Araya et al. 2005).

This region is associated with \h2O maser emission at 22 GHz, CH$_3$OH
maser emission at 6.7 GHz, and H$_2$CO maser emission at 8 GHz
(Miralles, Rodr{\'\i}guez \& Scalise 1994; Slysh et al. 1999; Beuther
et al. 2002c; Araya et al. 2005).  Dense gas traced by CS and
CH$_3$CN, as well as (sub)mm continuum emission is observed (Bronfman
et al. 1996; Sridharan et al. 2002; Beuther et al. 2002a; Williams et
al. 2004).  \nh3 (1,1) and (2,2) emission was detected first by
Miralles, Rodr{\'\i}guez \& Scalise (1994), and also by Molinari et
al. (1996) (source 83) and Sridharan et al. (2002) with single dish
telescopes.

Beuther et al. (2002b) report a CO outflow in the northwest-southeast
direction.  The geometric center of the
outflow, however, is about 10$''$ north of the 1.2 mm
emission peak.  Since the 1.2 mm continuum position
is consistent with that of the submm emission (Williams et al. 2004),
this offset is possibly caused by
pointing problems in the CO observations with the IRAM 30 telescope
(Beuther, H., private communication). The 1.2 mm emission shows an
extension of 15$''$ toward the northwest of the peak
emission. SiO J=2-1 emission is detected in the region with a
linewidth of 30 \kms-1\ at zero intensity (Beuther, H., private
communication).

The high resolution observations with the VLA and OVRO in this paper
reveal a collimated outflow in SiO and kinematics in the
dense core. In Section 2, we describe
details of observations. In Section 3, we present the main
observational results. In Section 4, we discuss the different
kinematic components in the region.  A summary is given in Section 5.

\section{Observations}
 
\subsection{VLA}

\subsubsection{ \NH3 observations}

The VLA\footnote{The National Radio Astronomy Observatory is operated
by Associated Universities, Inc., under cooperative agreement with the
National Science Foundation.} observations of IRAS 18566+0408 were
first conducted on 2001 July 23 in the \nh3 (J,K)=(1,1) and (2,2)
lines in the C configuration. To improve the S/N in the data,
follow-up observations were made from 2001 October
to 2003 January, in both CnB and DnC configurations in the \nh3 (1,1),
(2,2) and (3,3) inversion transitions.  The integration time on source
was typically less than 1 hour for each line.  The pointing center of the
observations was RA (2000) = $18^h59^m09^s.88$ and
DEC(2000)=$+04^\circ 12'13''.6$. We used  1849+005, 3C286 and
3C273 as the gain, flux, and bandpass
calibrators.  The detailed parameters of the
observations are summarized in Table 1.

The visibility data were calibrated using the NRAO Astronomical Image
Processing System (AIPS).  The uncertainty in the flux calibration is
about 10\%. The calibrated visibilities from different epochs were
combined for the same line and imaged in MIRIAD. The rms noise in the
(1,1), (2,2) and the (3,3) lines is about 2 mJy in a 3$''$ to 4$''$
synthesized beam per 0.6 \kms-1 wide channel.

\subsubsection{23 GHz and 43 GHz continuum}

The continuum observations at 23 GHz and 43 GHz were carried out with
the VLA on 2002 September 26 and 2003 February 04, respectively. At 43
GHz, we used the fast switching calibration scheme that alternated
between IRAS18566+0408 and the gain calibrator 1849+005 in a cycle of
2mins.  The total on-source time for IRAS18566+0408 was about 2 hours 
at 43 GHz, and 1
hour at 23 GHz.  Calibration and imaging were performed in
AIPS.  The flux calibration was done by comparing to 3C286.  The
absolute flux scales are accurate to about 10\%.  The rms is 0.1 mJy
in the 43 GHz image, and 0.14 mJy in the 23 GHz image, respectively.

\subsection{OVRO}

The OVRO observations of IRAS 18566+0408 were carried out during 2002
November to December. The SiO J=2-1 (v=0), HCN J=1-0 and \hcop J=1-0
lines were observed simultaneously in the lower sideband, along with
87 GHz continuum.  The SiO line was observed with a total bandwidth of
31 MHz and a spectral resolution of 0.5 MHz (1.7 \kms-1). The HCN line
was observed with a bandwidth of 30~MHz at a resolution of 1~MHz
(3.4~\kms-1). The \hcop\ line was observed with a bandwidth of 7.5~MHz
at a resolution of 0.25~MHz (0.8~\kms-1).  In addition, the analog
correlator provided continuum measurements of 4~GHz bandwidth.  The
pointing center of the OVRO observations was the same as that of the
VLA observations.  The detailed parameters of the observations are 
summarized in Table 1.

The visibility data were calibrated in the OVRO MMA package 
and exported to MIRIAD for imaging. The rms is
17 mJy per 1.7 \kms-1 channel for the line images, and
0.5 mJy for the continuum.
The \hcop\ emission is extended and suffers missing short spacing
fluxes, thus, is not presented  in this paper.

\section{Results}

\subsection{Continuum Emission}

No emission is detected at 23~GHz at a $1\sigma$ rms of 0.14~mJy.
Figure 1 presents images of continuum emission at 43~GHz (or 7.0~mm)
and 87~GHz (or 3.4~mm).  The 43~GHz emission shows a compact feature
with the peak position at RA (2000) = $18^h 59^m 09^s.99$, Dec (2000)
= $04^\circ 12' 15''.7$. The emission appears to be slightly resolved
with the $2''.7 \times 1''.3$ beam at a position angle of
$-7.6^\circ$. The peak and integrated flux densities are 1~mJy/beam
and 1.7~mJy, respectively, with a $1 \sigma$ error of 0.1~mJy. The
emission has an extension in the northeast-southwest direction, which
appears to be different from the position angle of the beam.  The
deconvolved size of the emission is $2''.0 \times 1''.2$ with a position
angle of $24^\circ$.  This corresponds to a size along the major axis
of $1.3 \times 10^4$~AU.

The 87 GHz emission is resolved with a $4''.9 \times 4''.5$ beam using
natural weighting.  The emission consists of a dominant peak, MM-1,
coincident with the peak of the 43GHz emission to better than $0''.1$,
and a secondary peak, MM-2, at RA (2000)
= $18^h 59^m 09^s.21$, Dec (2000) = $04^\circ 12' 22''.6$. The peak
flux density of MM-1 is 18 mJy/beam, with an integrated flux density
of 31 mJy. MM-2 is much weaker, with a peak flux density of 2.6
mJy/beam.  The uncertainty in these measurements is about 15\%.  There
appears to an extented filament at a position angle of
$-58^\circ$ connecting MM-1 and MM-2, which is 
better seen in the lower resolution (11$''$) 1.2mm map in Beuther et
al. (2002a).

Figure 2 shows the spectral energy distribution of the continuum peak
MM-1.  The 1.3~cm 3$\sigma$ upper limit ($1\sigma = 0.14$ mJy) is from
this paper. The 3.6~cm continuum detection is from Carral et
al. (1999). The 2~cm and 6~cm data are from Miralles et al. (1994) and
Araya et al. (2005). The 1.2~mm, 850~$\mu$m and 450~$\mu$m measurements
are from the IRAM 30-m telescope and JCMT (Beuther et al. 2002c;
Williams et al. 2004). The mid to far infrared data are from IRAS,
MSX and Spitzer IRAC measurements.

For emission at longer cm wavelengths, the contribution from dust is
negligible. Toward MM-1, a faint continuum source was detected at 2cm
(0.7 mJy) by Araya et al. (2005), and at 3.6 cm (0.7 mJy) by Carral et
al. (1999). However, we fail to detect the source at 1.3cm at an
angular resolution of $1''$ and a $1\sigma$ rms of 0.14 mJy. The
detections by Araya et al. (2005) and Carral et al. (1999) were made
at resolutions of a few arcseconds, and can be reconciled with the non
detection in this paper if the source is more extended than 1$''$.
However, an inconsistency remains at 2cm at which Miralles et
al. (1994) failed to detect the source with a $1\sigma$ rms of 0.16
mJy at a resolution of 5$''$. A possible reconciliation is that
the flux varies with time.  Despite the apparent differences, the
faintness of the cm emission indicates that the massive star in this
region is still extremely young in its evolution, and has not produced
significant free-free emission.

The measurements from wavelengths shortward of 1.2 mm have poorer
spatial resolution and sample a much larger area in the region.  We
fit a greybody model to the entire spectral energy distribution from
radio to infrared (IR) wavelengths. The far-IR measurements have a
typical resolution of $\sim 1'$. The overall spectral energy
distribution can be fitted by three dust components, with temperatures 
of 210, 58 and 30K, respectively. The total luminosity of the region determined
mainly by mid to far IR data at $\sim 1'$ resolution is $8 \times
10^4$ \lsun. The fluxes at the mm and submm wavelengths give a
spectral index $\alpha$ of 3.9, defined as $F_\nu \propto \nu^\alpha$,
or $\beta$ = 1.9.

For the compact continuum source MM-1, we use the high resolution 7~mm
and 3~mm data to derive a power law index more appropriate for a mass
estimate.  To minimize the difference in beam size between the two
frequencies, we image the 87~GHz data with a uniform weighting of the
visibilities and obtain a peak flux density of 10 mJy/beam with a
$3''.3 \times 2''.0$ beam. These two values (10 and 1.0~mJy/beam)
produce an upper limit to the spectral index $\alpha$ of 3.3, or an
upper limit to the emissivity index $\beta$ of 1.3.
Assuming that the dust reaches an equilibium with the gas through
collision at this high density environment (Burke \& Hollenbach 1983),
we approximate the dust temperature by the gas kinetic temperature of
80 K measured in \nh3 (see Sections 3.2, 3.3 and 4.1). For a dust
opacity law $\kappa \propto \nu^\beta$, and $\kappa(250\mu m) = 12$
cm$^2$g$^{-1}$ from Hildebrand (1983), we obtain a mass within the
$4''.9 \times 4''.5$ beam ($\sim$ 30,000 AU) of 70 \msun\ for $\beta =
1.3$.  This mass is a small fraction of the mass ($2 \times 10^3$
\msun) estimated from the 1.2mm emission for the entire region
(Beuther et al. 2002a).

For the continuum peak MM-2, the non detection at 7mm gives a
3$\sigma$ upper limit of 0.3~mJy. This value and the peak flux density
of 2.6 mJy/beam at 87~GHz yield a spectral index of $\gsim 3$.
Using assumptions similar to those for MM-1, we estimate the mass in
the MM-2 core. With a temperature of 30K derived from the \nh3
emission, we obtain a mass of 27 \msun\ for $\beta = 1.3$, and 13
\msun\ for $\beta = 1$.

\subsection{Line Emission}

Figure 3 presents the integrated emission of the \nh3 (J,K)=(1,1),
(2,2), (3,3) lines obtained from the VLA, and the SiO J=2-1 and HCN
J=1-0 transitions obtained from OVRO.  In the \nh3 (1,1) and (2,2)
lines, there appears to be extended emission in the
northwest-southeast direction over a scale of 40$''$. This component
is relatively cold as the (2,2) emission is less extended than the
(1,1) line. MM-2 is associated with \nh3 gas and lies in a molecular
ridge connecting MM-1 and MM-2 that is also seen in the dust emission.
In addition to the extended gas component, a compact \nh3 emission
component associated with MM-1 is seen in all three \nh3 lines, with a
deconvolved size of $1''.2$.  The peak intensity of the integrated
emission in the (2,2) line is 0.22 Jy~\kms-1/beam, similar to the
value in the (1,1) line. Thus the gas in this component is relatively
warm. In the \nh3 (3,3) line, there is also extended emission in the
northwest-southeast orientation.  The extension, at a position angle
of 135$^\circ$, is similar to the
emission in SiO and HCN. Unlike in the case of the \nh3 (1,1) and
(2,2) emission, MM-2 is located toward the edge of the \nh3 (3,3), SiO
and HCN emission.

To show detailed kinematics in the region, Figure 4 presents the channel maps 
of the \nh3 (J,K)=(1,1), (2,2) and (3,3) lines.
The extended emission is present mostly at velocities from 84 to 87 \kms-1, 
with a peak velocity of 85.2 \kms-1 corresponding to the cloud systemic velocity
(Bronfman et al. 1996).
The typical linewidth for the extended emission
is about $1-2$ \kms-1 in FWHM and the typical temperature is
$< 15$K (see the temperature map in Figure 6 and discussions in Section 4). 
The relatively narrow linewidth and low temperature in the gas indicate
that this extended component is from the quiescent gas in the core.

At velocities of 86 to 87 \kms-1, there appears to be a molecular ridge
(position angle of 148$^\circ$) between MM-1 and MM-2 in the (1,1) and (2,2) 
emission. MM-2 coincides
with a peak in the \nh3 emission (see channel 86.4 \kms-1).
At velocities less than 84 \kms-1 and greater
than 87 \kms-1, there appears to be compact \nh3 emission
toward the position of MM-1. This compact emission, with
a FWHM of 5.5 \kms-1 measured at $3''$ resolution,
is strong in the (2,2) emission relative to the (1,1) emission,
indicating that the \nh3 gas is rather warm. The ratio of the \nh3 (1,1)
and (2,2) lines gives a rotational temperature of 45 K.
 
In the \nh3 (3,3) line, the extended component seen in the (1,1) and
(2,2) is not as dominant. A compact emission component toward the
position of MM-1 stands out prominently. Since the (3,3) transition
has a higher upper energy level (124K) as compared to the (1,1) (23K)
and (2,2) (65K) lines, the (3,3) emission confirms that the compact
component is rather warm.

In addition to the compact component toward MM-1, there appear to be
four additional compact emission components in the \nh3 (3,3) line,
two toward the east of MM-1 in the velocity channels of 84.6 and 85.5
\kms-1, one to the west of MM-1 in the velocity channel of 85.2
\kms-1, and one to the northwest of MM-1 from velocities of 85.5 to
85.8 \kms-1.  We refer to these features as `A', `B', `C' and `D',
respectively (The crosses in the channel maps in Figure 4). There
appear to be no corresponding emission peaks in the \nh3 (1,1) and
(2,2) lines.  Unlike the broad \nh3 (3,3) line emission toward MM-1,
these four components have rather narrow velocity width of about 1.5
\kms-1 in FWHM, but extended line wing emission. We will discuss these
features further in Section 4.

Besides the compact emission components in Figure 4c, the \nh3 (3,3)
emission also shows an extended structure in the southeast-northwest
direction. This structure has rather broad line wings (15 \kms-1 from
the cloud velocity), and high temperatures of 70 K. MM-2 is not
associated with any peaks of the (3,3) emission.

Figure 5a presents the channel maps of the SiO J=2-1 transition.  The
SiO emission is elongated and lies mostly to the northwest of MM-1 at
a position angle of 135$^\circ$, similar to the extended emission in
the \nh3 (3,3) line.  The SiO emission is present from velocities of
70 to 93 \kms-1.  There appears to be higher velocity SiO emission
toward MM-1, but none toward MM-2. Since SiO abundance is typically
low in quiescent clouds (Ziurys, Griberg \& Irvine 1989) and is
enhanced by a few orders of magnitude in outflows (e.g. Zhang et
al. 1995) due to shock processes (Pineau des For\^ets, Flower \&
Chi\`eze 1997), the SiO emission here most likely traces a well
collimated outflow originated from MM-1.  The orientation of the SiO
outflow is consistent with the bipolar CO outflow reported by Beuther
et al. (2002).

Figure 5b presents channel maps of the HCN 1-0 emission. 
The HCN 1-0 transition has three hyperfine components (F=1-1, 2-1 and
0-1) at relative frequencies corresponding to 4.8, 0 and -7.1 \kms-1,
respectively. We set the hyperfine component F=2-1 at a
\vlsr\ of 85.2 \kms-1, the cloud systemic velocity. It appears that the
emission from the F=1-1 component is weak. The F=2-1 and 0-1
components are detected around 80 and 73 \kms-1, respectively (see also Figure 
7), 5 \kms-1\ blue shifted from the cloud velocity.
The HCN emission arises mainly in two strong peaks: One is associated with 
the dust peak MM-1, the other is 4$''$ offset from MM-2 and
coincides with the SiO peak in the outflow.
Little emission is detected toward MM-2.

The measured flux ratios of the three HCN hyperfine components
(F=0-1, 2-1 and 1-1) amount to 1:1:0.2. Under the LTE condition, line
ratios vary from 1:5:3 for optically thin emission to 1:1:1 for
optically thick emission. The measured ratios are not consistent with
partially optically thick gas under LTE. Anomalous hyperfine ratios of
HCN have been found in both cold dark clouds and warm clouds around
\hii\ regions (Walmsley et al. 1982; Cernicharo et al. 1984).
Possible causes involve overlapping hyperfine components of higher
rotational transitions or core-envelope density structures/velocity
gradients in the cloud (Gonzalez-Alfonso \& Cernicharo 1993). The
ratios measured in IRAS 18566+0408 are different from those in dark or
warm clouds. Missing short spacing flux in the interferometer
data can affect the observed core and envelope emissions differently,
which in turn affects the hyperfine ratios. Therefore, we do not
further investigate this issue quantitatively.

\subsection{Rotational Temperature}

We derive rotational temperatures of the \nh3 gas.  In the
calculation, we assumed LTE conditions in the gas and followed the
procedure outlined in Ho \& Townes (1983).  Figure 6 presents a map of
rotational temperature derived from the \nh3 (1,1) and (2,2) lines. In
most of the core, temperatures are around 10 K to 15 K. Higher
rotational temperatures of about 45K are found toward MM-1, and along
the ridge of SiO emission. In this high temperature region, there
exists an area where the rotational temperatures cannot be
derived. This is because the ratio of the (1,1) and (2,2) lines is
sensitive to temperatures only up to 50K (Ho \& Townes 1983). At
temperatures over 50 K, the ratio of the two lines approaches 1 for
the optically thick case, and 1.3 in the optically thin case for a
wide range of temperatures. A small error in the flux measurement will
result in a large uncertainty in rotational temperatures.  Thus, 
the blanked area along the SiO outflow in Figure 6 has even higher
temperatures. Assuming the same abundance for the ortho and para \nh3
species, we use the (3,3) and the (1,1) lines to obtain a temperature
estimate of 70 K for blanked area in the outflow region.

\section{Discussions}

\subsection{Nature of the Continuum Peaks}

The 3~mm continuum emission reveals two peaks, MM-1 and MM-2, bridged
by a faint extended filament. The dominant peak MM-1 coincides with
the compact 7mm continuum source, faint cm continuum emission (Araya
et al. 2005; Carral et al. 1999), and the strong peaks in \nh3, SiO
and HCN emission. The rotational temperature estimated from \nh3 is 45
K, corresponding to a kinetic temperature of 80 K (Danby et al. 1988).
The luminosity of the internal source required to produce the heating
can be estimated by the following equation (Scoville \& Kwan 1976)
$$ T_D = 65 ({0.1pc\over
r})^{2/(4+\beta)} ({L_{star}\over 10^5 L_\odot}) ^{1/(1+\beta)} ({0.1\over
f})^{1/(4+ \beta)} K.$$ 
Here $\beta$ is the power law index of the dust
emissivity at far infrared wavelengths, $f$ = 0.08 cm$^2$~g$^{-1}$ is the
value of the dust emissivity at 50$\mu$m, and $r$ the core radius.
If the dust and gas reach a thermal equilibrium in the high density
environment, $i.e., T_D$ = 80K at a radius $r$ = 7000 AU, and $\beta = 1$,
we estimate the luminosity of the embedded source to be
$6 \times 10^4$ \lsun. This value is in rough agreement with the far IR luminosity
observed for the region. The high luminosity, strong dust and molecular
line emission, and high temperature in the gas
all indicate embedded massive protostar(s) toward MM-1.

The secondary mm peak MM-2 lies in a molecular ridge seen in \nh3 and
coincides with a local \nh3 peak in the (1,1) and (2,2) transitions
(see channels 85.7 and 86.4 \kms-1\ in Figures 4a and 4b). This
molecular ridge appears to correspond to the dust emission seen at 3mm
and 1.2mm.  \nh3 gas temperature toward MM-2 is about 30 K. The large
amount of dense molecular gas and a local peak in the gas and dust
emission may indicate embedded protostar(s) in the core. If the
heating of the gas and dust is due to an internal source, we find a
luminosity of $8 \times 10^3$ \lsun\ for the protostar, through a
similar analysis as described above for MM-1. On the other hand, if
the heating is partially due to the molecular outflow in the region
(see discussions below), the total luminosity for MM-2 would be lower.

\subsection{Massive Molecular Outflow in SiO and \nh3}
\label{outflowsection}

The SiO emission delineates a bipolar molecular outflow in the
region. Figure 7 shows the position-velocity plots of the SiO, HCN,
and \nh3 (2,2) and (3,3) emissions along the major axis of the SiO
emission at a position angle of 135$^\circ$.  Toward the northwest of
MM-1, the SiO emission is blue shifted with respect to the cloud
systemic velocity of 85.2 \kms-1.  The terminal velocity of the
blue-shifted SiO emission is about 15 \kms-1\ from the cloud velocity.
Close to the peak MM-1, both the blue- and red-shifted SiO emissions
are detected up to 30 \kms-1 (3 $\sigma$ level) from the cloud
velocity. The blue-shifted SiO emission extends 15$''$ to the
northwest. The red-shifted emission is far more compact spatially,
with a peak detected only 2$''$ southeast of MM-1.

As shown in Figure 7, there exists cold and quiescent \nh3 gas along
the outflow direction. The \nh3 emission peaks at \vlsr\ of 85.2
\kms-1, and has a narrow FWHM of $< 2$ \kms-1.  In addition to the
cold gas, there exists blue-shifted high velocity emission up to
\vlsr\ of 70 \kms-1, 15 \kms-1\ from the cloud core velocity. This
high velocity gas, offset to the northwest from the continuum peak, is
part of the blue-shifted molecular outflow. The gas has an estimated
temperature of 70 K.  Although \nh3 is a reliable tracer of dense gas
in molecular cloud cores, it can be affected by molecular outflows
associated with both low and high mass stars (L1157: Tafalla \&
Bachiller 1995; IRAS20126+4104: Zhang et al. 1999). The high velocity
\nh3 and SiO gas has been likely accelerated and heated by shock
processes in the outflow.

Although the major axis of the SiO outflow agrees with that of the CO
2-1 outflow obtained at 11$''$ resolution (Beuther et al. 2002b), the
CO outflow exhibits nearly symmetric bipolar morphology with the
southeastern lobe much stronger than that in the SiO.  Furthermore,
the polarity of the SiO outflow appears to be the opposite of that of
the CO: The blue-shifted SiO emission lies to the northwest of the
star, while the blue-shifted CO emission lies to the southeast of the
star. This change of polarity between different tracers has been seen
toward other objects (e.g. IRAS 20126+4104; Cesaroni et al.  1997;
1999). One plausible explanation is that the outflow axis lies almost
in the plane of the sky and precesses. The low density CO gas traces
the wide angle component in the outflow, while the high density SiO
gas traces the well collimated jet component in the outflow.  On the
other hand, CO outflows toward massive star forming regions are often
unresolved by single dish telescopes, and break into multiple bipolar
outflows at high angular resolution (e.g.  I05358, Beuther et
al. 2002; AFGL 5142, Zhang et al 2007). Furthermore, the SiO and CO
may trace different outflows as shown in AFGL~5142 (Zhang et al.
2007; Hunter et al. 1999) and the Orion South region (Zapata et al. 2006).  
High resolution CO images of the outflow will help resolving the
difference.

We compute the mass, momentum and energy in the SiO outflow.  
Using an SiO to \h2\ fractional abundance of $10^{-7}$ (Zhang et al
1995), an excitation temperature of 70K derived from \nh3, and
assuming optically thin SiO emission, we obtain outflow mass, momentum and
energy of 18 \msun, 200 \msun~\kms-1, and $4.0 \times 10^{46}$ ergs,
respectively, in the blue-shifted lobe. Likewise, we obtain 9 \msun,
70 \msun~\kms-1, and $1.0 \times 10^{46}$ ergs in the red-shifted
lobe. Despite the uncertainty in the SiO abundance, the total mass,
momentum and energy of 27 \msun, 270 \msun~\kms-1, and $5.0 \times
10^{46}$ ergs are in a rough agreement (within a factor of 2) with the
estimates from the CO outflow (Beuther et al. 2002b). 

The terminal velocity of 15 \kms-1\ and the length of the SiO outflow
(15$''$) yield a dynamical time scale ($T_{dyn}$ ) of $1.7 \times
10^4$ years. This value is a few times smaller than that of the CO
based on the lower angular resolution data (Beuther et
al. 2002b). Assuming momentum conservation between the outflow and the
underlying wind that powers the outflow: ${m}_{w}V_{w} =
{P}_{outflow}$, we can estimate the mass loss rate in the wind over
the dynamical time scale of the outflow. The wind velocity $V_{w}$ can
vary from 100 \kms-1\ in low-mass stars to 500 \kms-1\ in high-mass
stars (Zhang et al. 2005). Since the effect of inclination angle of
the outflow is not corrected, we use a lower value of 100 \kms-1\ for
the wind velocity. This gives a mass loss rate ($m_w/T_{dyn}$) of $1.5
\times 10^{-4}$ \msun yr$^{-1}$, and thus a lower limit to the mass accretion
rate of $1.5 \times 10^{-4}$ \msun yr$^{-1}$, since some material
presumably goes into the central protostar (Churchwell 2002).


\subsection{Heating and Weak Maser Emission}

The outflow apparently causes significant heating in the molecular
gas, especially toward the blue-shifted lobe. As shown in Figure 6,
the rotational temperature derived from the \nh3 (1,1) and (2,2) lines
is 45 K toward the position of MM-1.  However, the temperature in the
outflowing gas toward the blue-shifted lobe is higher, with values of
50-70 K.

The effect of heating is further demonstrated in Figure 8. We compute
ratios of the \nh3 (3,3) and (1,1) emission. As the (3,3) transition
has an energy level of 124 K, the higher ratios in general represent
higher gas temperatures for thermal emission. As shown in Figure 8,
the \nh3 gas in the outflow region exhibits consistently higher line
ratios, as expected from high temperature regions.  Toward the
positions `C' and `D', there appears to be heated \nh3 gas in a bow
shape. The \nh3 spectrum toward `C' shows red-shifted line wings,
while the \nh3 spectra toward `D' show blue-shifted line wings.
Furthermore, the tips of the bows point away from each other, as seen
in the upper-right panel in Figure 8.  It is possible that the heating
in \nh3 traces another outflow, which is not seen in SiO. The heating
likely arises from bow shocks as high velocity gas impinges on the
cloud core. The potential driving source should lie in between `C' and
`D'. However, no dust continuum emission is detected at a 3$\sigma$
limit of 4 \msun.

Figure 9 presents \nh3 spectra toward positions `A', `B', `C' and `D'.
All \nh3 spectra display line wing emission 15 \kms-1 blue shifted
from the cloud velocity. Toward `A', `B' and `C', the (3,3) line has a
FWHM of $<$ 1.5 \kms-1, smaller than the $\sim$ 3 \kms-1\ FWHM in the
(1,1) and (2,2) lines.  The compact morphology, the narrow linewidth,
and a lack of corresponding peaks in the \nh3 (1,1) and (2,2) emission
indicate that the compact (3,3) emission arises from population
inversion, similar to maser emission.  Maser inversion of the \nh3
(3,3) has been detected toward a number of sources (e.g. W51, Zhang \&
Ho 1995, NGC6334, Kraemer \& Jackson 1995, Beuther et al. 2007; DR
21(OH), Mangum \& Wootten 1994; Mauersberger, Wilson \& Henkel 1986;
IRAS 20126+4104, Zhang et al. 2001).  \nh3 (3,3) inversion can form
through collisional excitation of \nh3 by \h2 (Walmsley \& Ungerechts
1983).  Through collisions with \h2, the upper level of the \nh3 (3,3)
(denoted as \nh3 (3,3)$^+$) exchanges with its (0,0) state while the
lower level of the (3,3) exchanges with the (1,0). Since the
transition between the (3,3)$^+$ and (0,0) involves a change of parity
and thus is more preferred, the (3,3)$^+$ state can be overpopulated.

\nh3 (3,3) masers are often observed in outflows. In the cases of IRAS
20126+4104 and NGC 6334 (Kraemer \& Jackson 1995; Zhang et al. 2001),
(3,3) masers are detected in the vicinity of bow shocks where outflow
wind interacts with the cloud gas. The high velocity \nh3 gas detected
toward `A', `B' and `C' (see Figure 9) suggest a similar
scenario. However, the spatially compact \nh3 (3,3) emission appears
to be resolved at an resolution of $1''$, suggesting that the emission
is not strongly amplified.

\subsection{Kinematics in the \nh3 Core}

In MM-1, the compact \nh3 emission with a Gaussian-like profile and
large linewidth distinguishes itself from the relatively smooth \nh3
emission in the cloud core.  The \nh3 emission toward this position
has much broader linewidth: 5.5 \kms-1 at a spatial resolution of
3$''$. We image the visibility data from the VLA C array only and
obtain an angular resolution of 1$''$.  From this image, we obtain a
fitted FWHM of 8.7 \kms-1 in the (3,3) line. This increase indicates
additional broadening in the spectral lines towards the inner part of
the \nh3 core.  The \nh3 compact structure has a size of 1.2$''$ or
8000 AU.  We estimate the mass in this compact structure following Ho
\& Townes (1983).  With a rotational temperature of 45 K, size of
1.2$''$ and $[{NH_3 \over H_2}] = 10^{-7}$ (Harju et al. 1993), and
the assumption of LTE, we obtain a mass of 60 \msun. This value is
consistent with the mass estimated from the dust emission.

The broad \nh3 linewidth toward MM-1 can arise from outflow,
infall/rotation or relative motion of multiple objects unresolved
within the synthesized beam.  A collimated molecular outflow is
present in the SiO emission with high velocity emission toward the
northwest and the southest of MM-1. The effect of the outflow also
appears in the \nh3 emission, especially toward the northwest of MM-1
along the outflow lobe. Can the molecular outflow produce the line
broadening seen in \nh3 toward MM-1?  The SiO emission is shifted from
the cloud velocity.  On the contrary, the \nh3 emission peaks mostly
at the cloud velocity and appears to be Gaussian in
profile. Furthermore, the mass and momentum in the outflow within the
$5''.0 \times 4''.7$ area (the synthesized beam of the SiO data) of
MM-1 are 1.5 \msun\ and 10 \msun~\kms-1, respectively. Similarly, we
compute the same quantities from the \nh3 gas over a scale of $1''.2$,
and find the mass and momentum of 60 \msun\ and 250 \msun~\kms-1,
respectively. The fractional abundances of \nh3 and SiO may be
uncertain and thus can affect the estimates provided
above. Nevertheless, the comparison between the masses in SiO and \nh3
shows that toward the most central region of the core only $<$ 3\% of
the material traced by the \nh3 emission is from the molecular
outflow. Since the outflow mass is calculated over the area ($5''.0
\times 4''.7$) 15 times larger than that of the \nh3 emission, the
actual contribution from the outflow can be even smaller.  Thus, it is
unlikely that the molecular outflow is the main contributor to the
\nh3 linewidth.

The remaining possibilities for the large \nh3 linewidth are motions
such as infall/rotation or relative motion of multiple cores within
the synthesized beam.  Higher angular resolution observations in dust
continuum and spectral lines will be fruitful in distinguishing these
possibilities.  If the \nh3 linewidths are due to rotation and infall,
similar to the signature seen in \nh3 toward IRAS 20126+4104 (Zhang et
al. 1998), the dynamical mass, assuming gravitationally bound motion,
derived using $M = {V_{rot}^2 R \over 2 G}, $ is 35 \msun, for
$V_{rot}$ = 3 \kms-1\ at R = 7000AU.  This is compatible with the mass
estimate from dust emission and \nh3 at a similar scale. The mass
infall rate, estimated using $4\pi R^2 n_{H_2} V_{infall}$, is $1.5
\times 10^{-3}$ \msun yr$^{-1}$ for $V_{infall}$ = 3 \kms-1\ and
$n_{H_2} = 10^{6}$ \cm3. Thus, the mass loss rate in the wind, $1.5
\times 10^{-4}$ \msun yr$^{-1}$ (Section \ref{outflowsection}), is
10\% of the infall rate.  Assuming that 10\% to 30\% of the infalling
mass is ejected in the outflow, the accretion luminosity, estimated
from ${G M \dot{M} \over r}, $ amounts to $3 - 4 \times 10^4$ \lsun,
about half of the far-IR luminosity.

\section {Conclusion}

We conducted observations of the high-mass protostellar candidate
IRAS 18566+0408 with the VLA and OVRO interferometers.

(1) We resolve a collimated outflow in SiO (2-1) emission. The outflow
is also detected in the line wings of \nh3 inversion transitions
and produces significant heating of the molecular gas (up to
70K). Compact features in the \nh3 (3,3) line are detected along the
outflow. The narrow linewidths of $< 1.5$ \kms-1\ suggest that they
are weakly amplified maser emission.

(2) The 87 GHz emission reveals two peaks MM-1 and MM-2. The internal heating
in the MM-1 core calls for embedded massive young star(s). The \nh3 linewidth
toward MM-1 is much broader than the typical linewidth
of $<$ 2 \kms-1 in the extended core, and  increases
inward from 5.5 \kms-1 at the 3$''$ scale to 8.7 \kms-1 at the 1$''$ scale.
The motion is consistent with rotation/infall, but can also arise
from relative motions of unresolved protostellar cores.

\begin{acknowledgements}
 {We appreciate Editor M. Walmsley for his valuable comments.
H. B. acknowledges financial support by the Emmy-Noether-Programm of the
Deutsche Forschungsgemeinschaft (DFG, grant BE2578). Y. C. thanks
the support by the NSFC Grant 110133020.}
\end{acknowledgements}

\clearpage
 
\begin{table*}
\begin{center}
\caption{List of Observational Parameters}
\label{tab:3.1}
\begin{tabular}{l|lllll}
\hline
Instrument & Date of      & Line        &  Bandwidth & Spectral   & Integration  \\
           & Observations &             &     (MHz)  & Res. (\kms-1) &  Time (hr) \\
\hline
VLA-CnB    & 2001/07/23   & \nh3\ (1,1),(2,2) &   3.12     &  0.6       &  0.7  \\
VLA-DnC    & 2001/10/01   & \nh3\ (3,3)       &   3.12     &  0.3       &  1.0  \\
VLA-CnB    & 2002/09/26   & \nh3\ (1,1),(2,2) &   3.12     &  0.6       &  1.2     \\
VLA-CnB    & 2002/09/26   & \nh3\ (3,3)       &   3.12     &  0.6       &   1.0    \\
VLA-CnB    & 2002/09/26   & 23 GHz      &   25       &  -         &    1.0   \\
VLA-DnC    & 2003/01/16   & \nh3\ (1,1),(2,2) &   3.12     &  0.3      &   0.6    \\
VLA-DnC    & 2003/01/24   & \nh3\ (3,3)       &   3.12     &  0.6      &   1.0    \\
VLA-DnC    & 2003/02/04   & 43 GHz      &   50       &  -        &   2    \\ \hline
OVRO-E     & 2002/11/02   & HCN (1,0)         &   30     &  3.4      &   4.0    \\
OVRO-E     & 2002/11/02   & SiO (2-1)         &   30     &  1.7      &   4.0    \\
OVRO-E     & 2002/11/02   & HCO$^+$ (1-0)     &   7.5     &  0.84      &   4.0    \\
OVRO-E     & 2002/11/02   & 87 GHz     &   4000     &  -      &   4.0    \\
OVRO-H     & 2002/12/01   & 87 GHz     &   4000    &   -        &   2.0    \\
\hline   
\end{tabular}
\end{center}
\end{table*}

\begin{figure*}
\includegraphics[width=17cm]{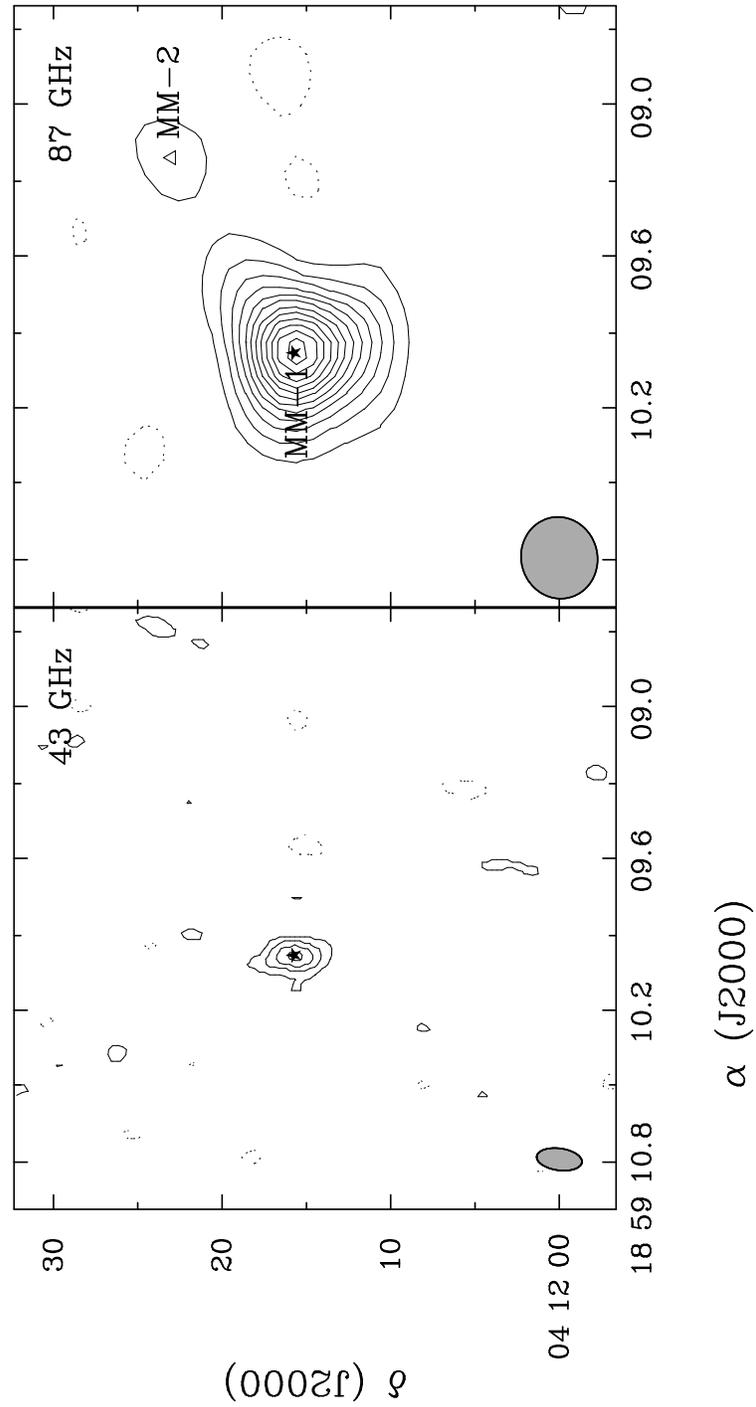}
\caption{Continuum emission at 43 GHz and 87 GHz toward IRAS 18566+0408.
The contour levels are in steps of 0.25 mJy/beam for the 43 GHz continuum image,
and 1.5 mJy/beam for the 87 GHz continuum image. The `star' symbol and
`triangle' mark the continuum peaks MM-1 and MM-2, respectively.
The size of the synthesized beam
is marked by the shaded ellipse at the lower-left corner of each panel.}
\end{figure*}

\begin{figure*}
\includegraphics[width=17cm]{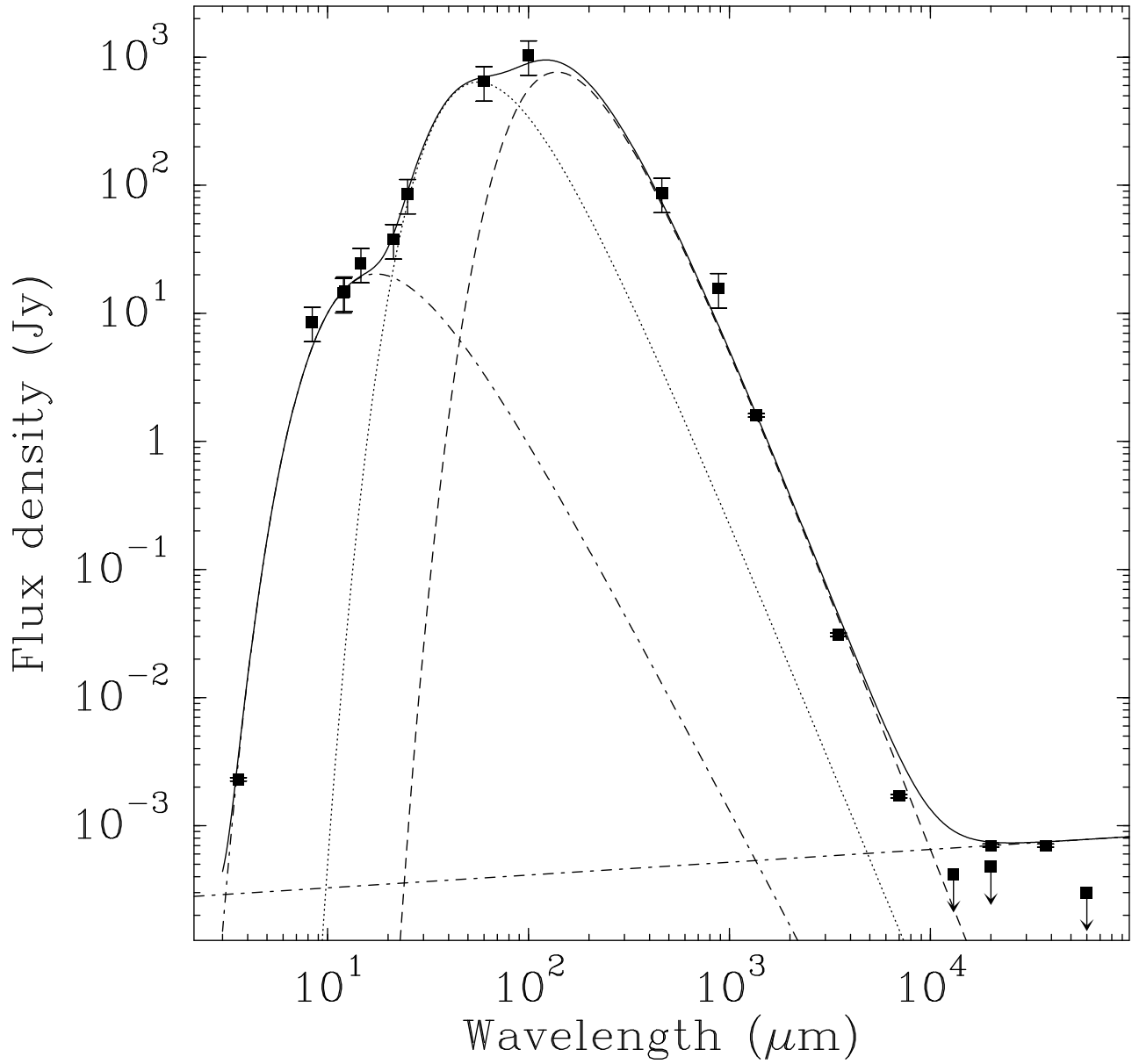}
\caption{The spectral energy distribution of IRAS
18566+0408. The dashed line represents a cold dust component, the
dotted line represents the warm dust component, the dash-dot line 
represents the free-free emission, and the solid line represents the
sum of all three components.}
\end{figure*}

\begin{figure*}
\includegraphics[width=17cm]{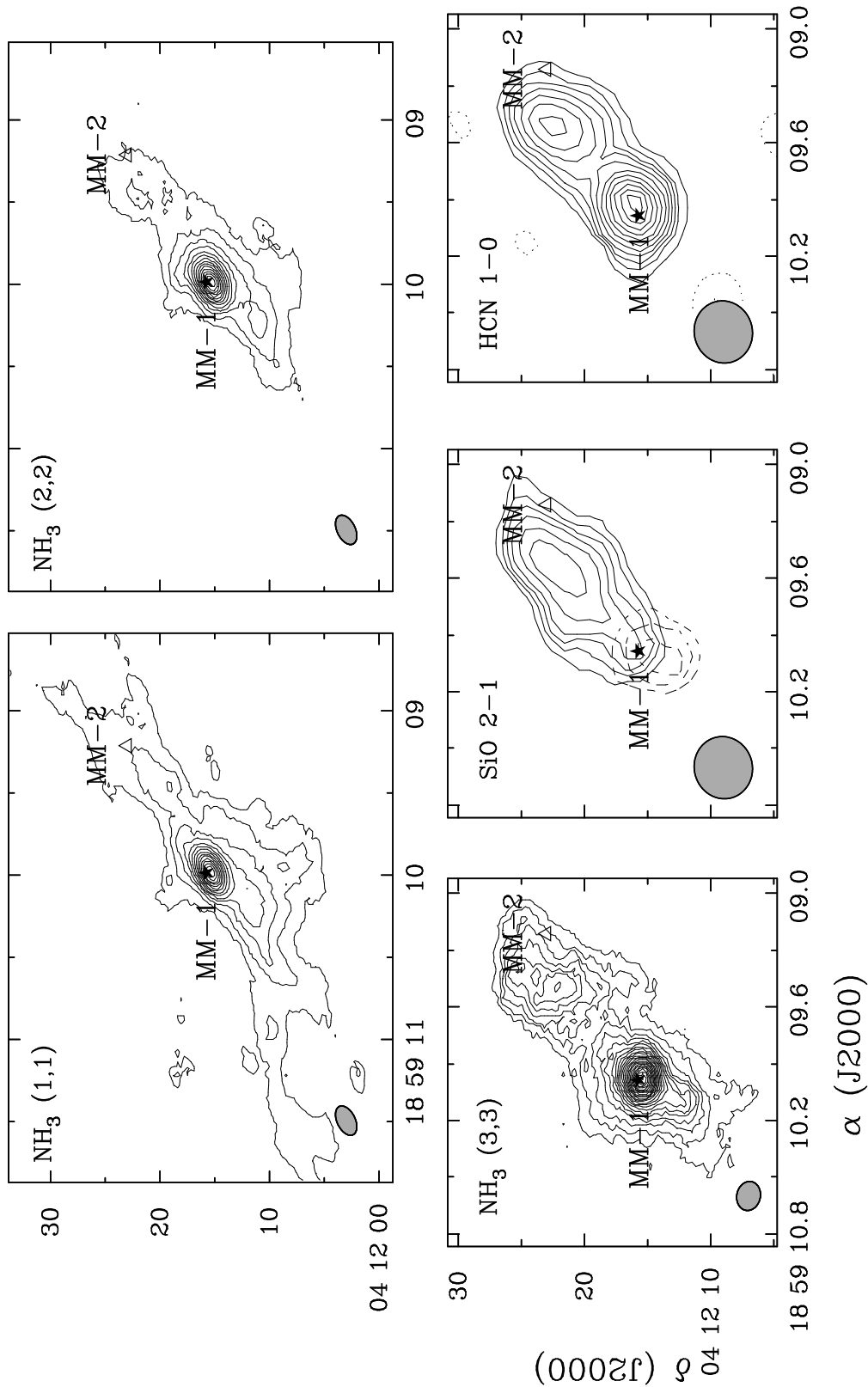}
\caption{The integrated emission of the \nh3 (J,K)=(1,1), (2,2), (3,3),
SiO J=(2-1), HCN J=(2-1) lines. The velocity range of the integration is
81 to 88 \kms-1\ for the \nh3 (1,1) and (2,2) lines. The \nh3 (3,3)
and HCN lines are integrated over the entire spectral line.
For the SiO line, the blue-shifted emission (solid contours) is integrated
from 50 to 83 \kms-1\, and the red-shifted emission (dashed contours) 
is integrated from 88 to 110 \kms-1.
The \nh3 images are made from the VLA-D and C configuration data. 
The SiO and HCN data images are made from the OVRO E configuration data.
The contour levels
are in steps of 0.015 Jy~\kms-1 for the \nh3 (1,1) and (2,2) lines,
0.02 Jy~\kms-1 for the \nh3 (3,3), 0.3 Jy~\kms-1 for the SiO (2-1), and
0.6 Jy~\kms-1 for the HCN (1-0).
The `star' symbol and
`triangle' mark the continuum peaks MM-1 and MM-2, respectively.
The beam sizes are plotted at the lower-left corner of each panel.}
\end{figure*}

\renewcommand{\thefigure}{\arabic{figure}\alph{subfig}}
\setcounter{subfig}{1}
\begin{figure*}
\includegraphics[width=17cm]{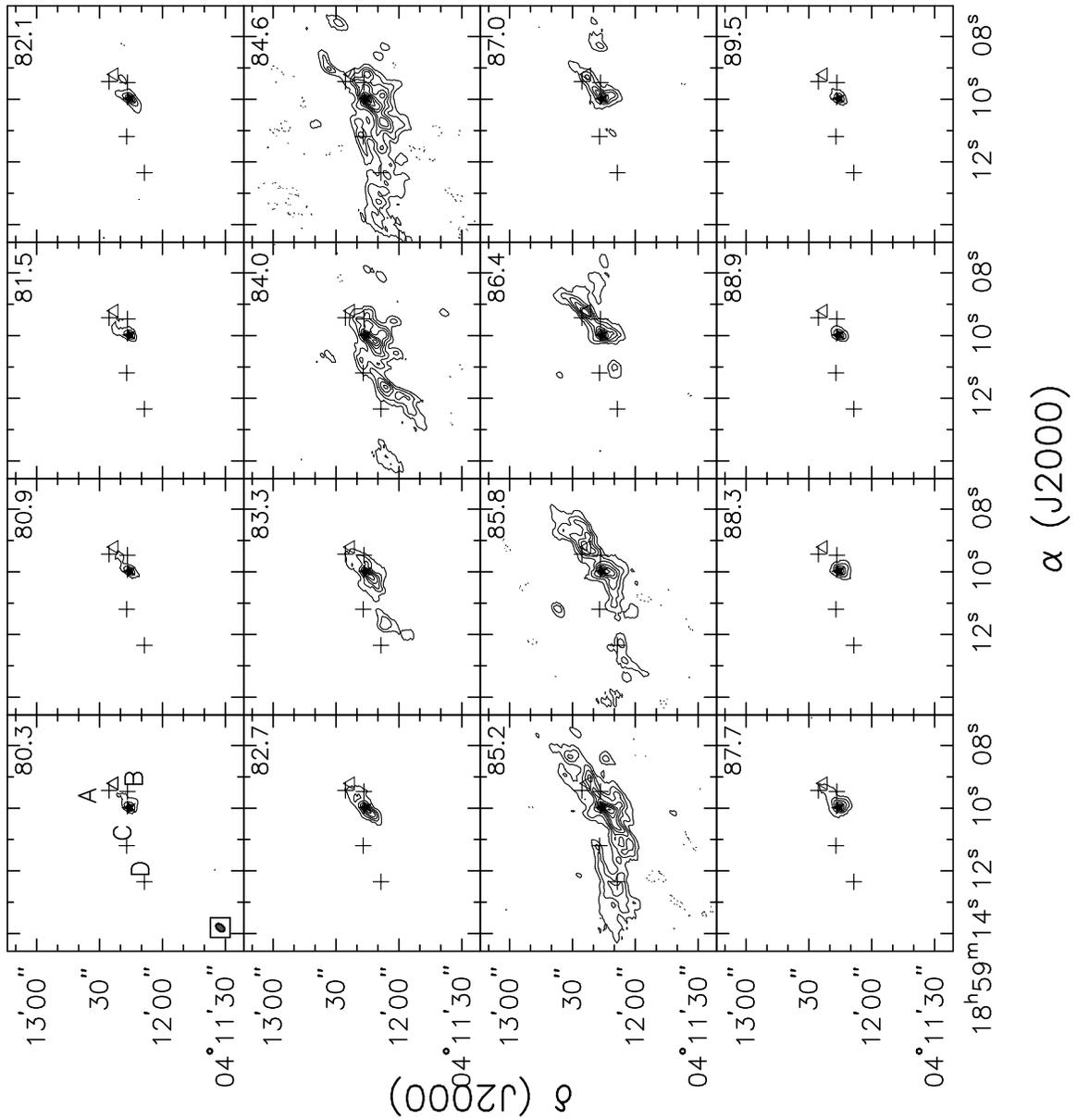}
\label{4a}
\caption{The channel maps of the \nh3 (J,K)=(1,1) (Fig. 4a), 
(2,2) (Fig. 4b)
and (3,3) (Fig. 4c) lines. The images were made using the VLA-D configuration
data only. The contour levels are in steps of 8 mJy/beam ($\sim 4 \sigma$)
for the \nh3 (1,1) and (2,2) lines and in steps of 9 mJy/beam ($\sim 3 \sigma$) 
for the (3,3) line.
The beam sizes are plotted at the lower-left of the first panel. The velocity
of the channel is plotted at the upper-right of each panel. The (1,1) and (2,2)
images have a spectral resolution of 0.6 \kms-1, while the (3,3)
images have a spectral resolution of 0.3 \kms-1.
The `star' symbol and
`triangle' mark the continuum peaks MM-1 and MM-2, respectively.
The crosses mark the position of the hot \nh3 (3,3) features.}
\end{figure*}

\addtocounter{figure}{-1}
\addtocounter{subfig}{1}
\begin{figure*}
\label{4b}
\includegraphics[width=17cm]{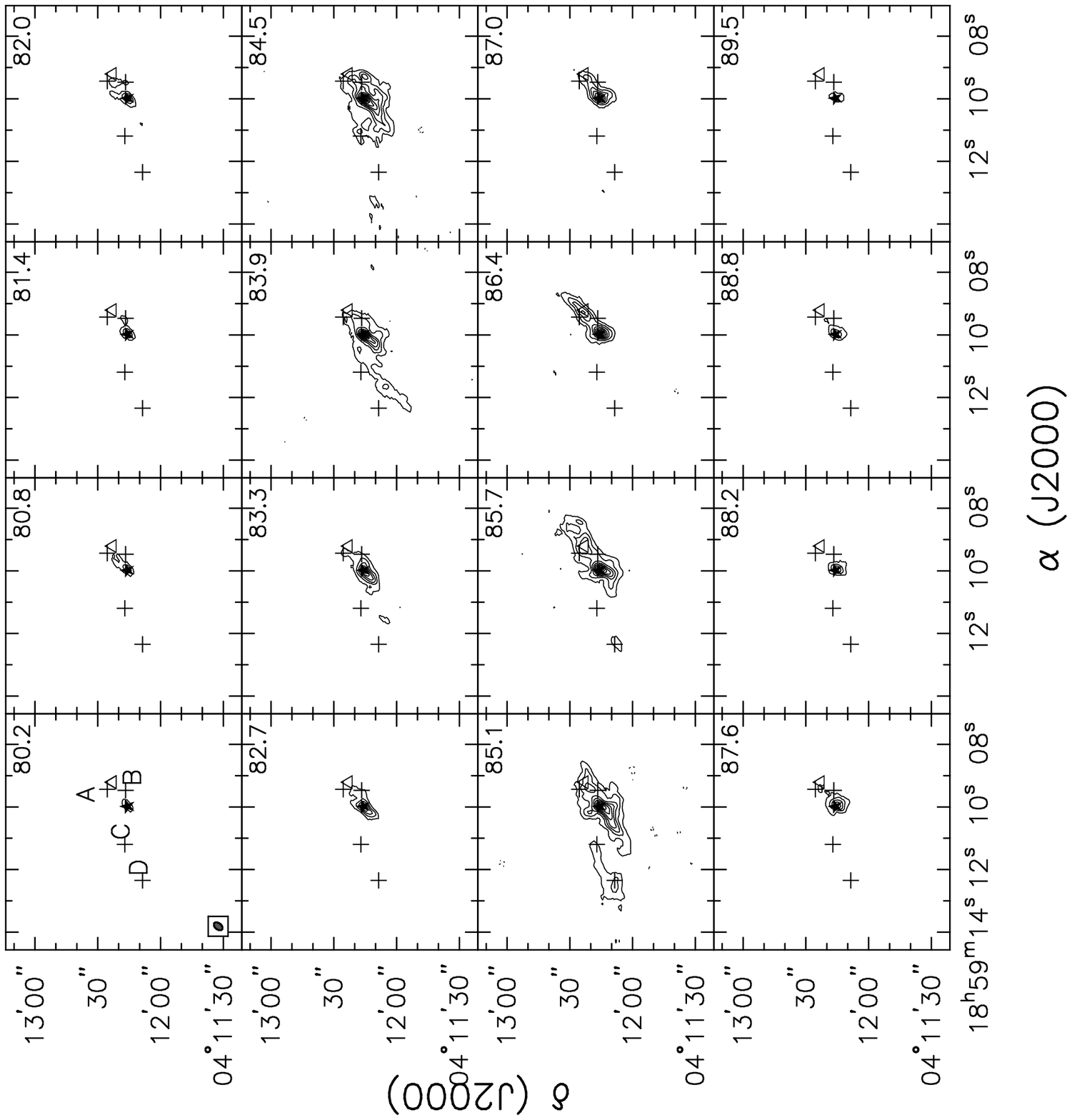}
\caption{}
\end{figure*}

\addtocounter{figure}{-1}
\addtocounter{subfig}{1}
\begin{figure*}
\label{4c}
\includegraphics[width=17cm]{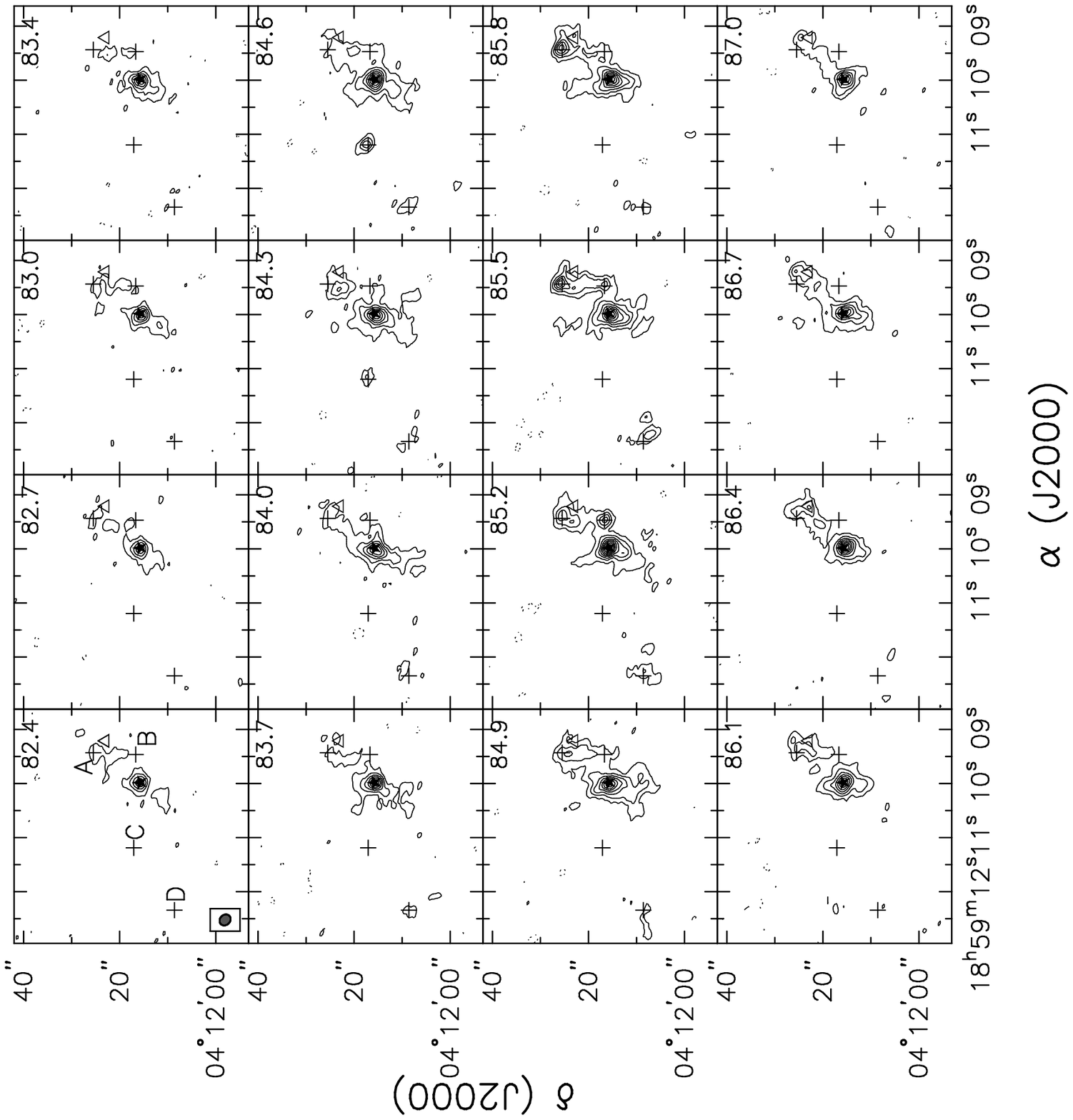}
\caption{}
\end{figure*}

\setcounter{subfig}{1}
\begin{figure*}
\label{5a}
\includegraphics[width=17cm]{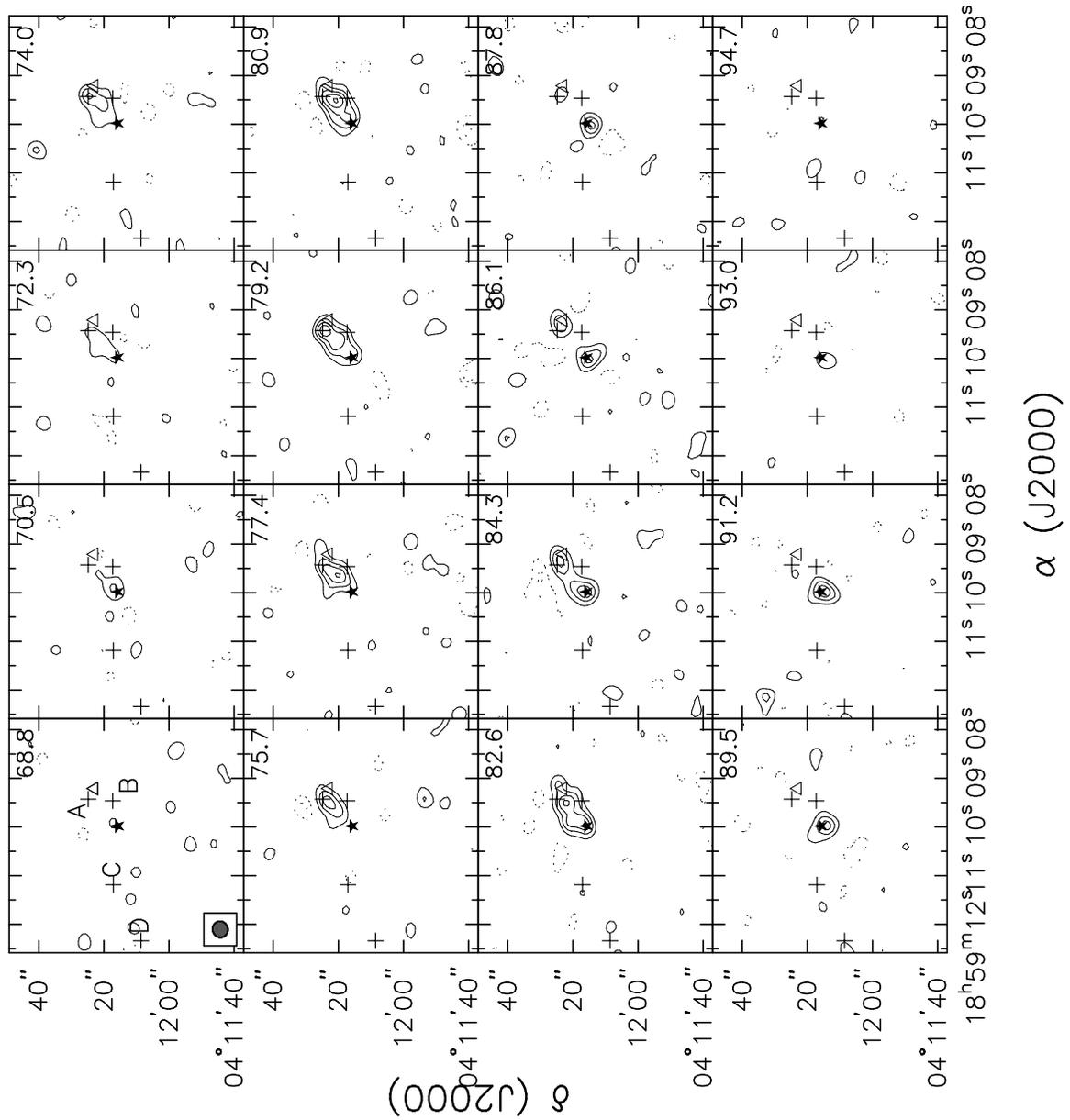}
\caption{The channel maps of the SiO J=(2-1) (Fig. 5a) and
HCN J=(1-0) (Fig. 5b) lines. The images were made using the OVRO E 
configuration data only. The contour levels are in steps of 0.05 
Jy/beam for the SiO line, and 0.07 Jy/beam for the HCN line ($\sim 3 \sigma$).
The beam sizes are plotted at the lower-left of the first panel. The velocity
of the channel is plotted at the upper-right of each panel.
The `star' symbol and
`triangle' mark the continuum peaks MM-1 and MM-2, respectively.
The crosses  mark the position of the hot \nh3 (3,3) features.}
\end{figure*}

\addtocounter{figure}{-1}
\addtocounter{subfig}{1}
\begin{figure*}
\label{5b}
\includegraphics[width=17cm]{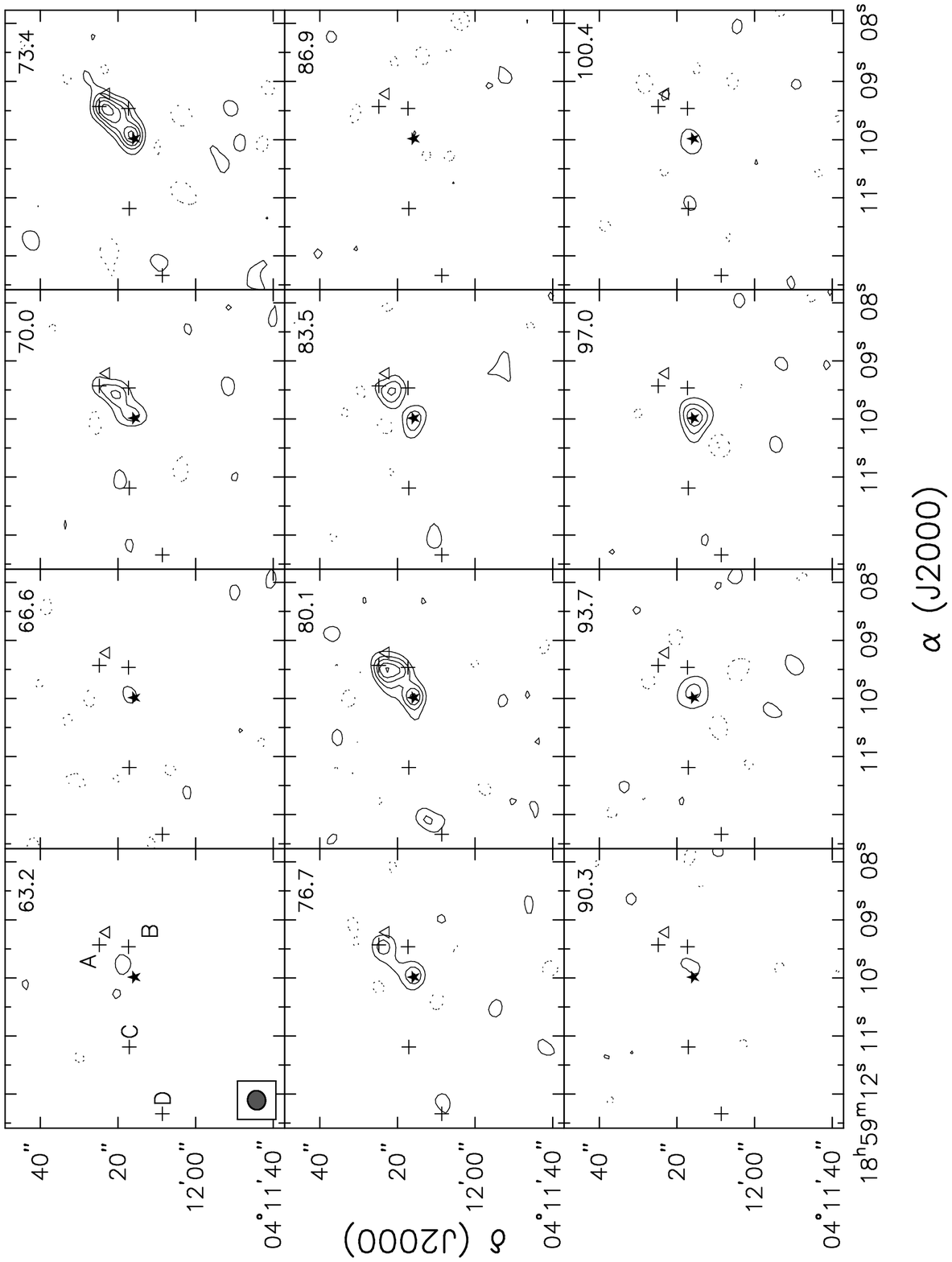}
\caption{}
\end{figure*}

\renewcommand{\thefigure}{\arabic{figure}}

\begin{figure*}
\label{6}
\includegraphics[width=17cm]{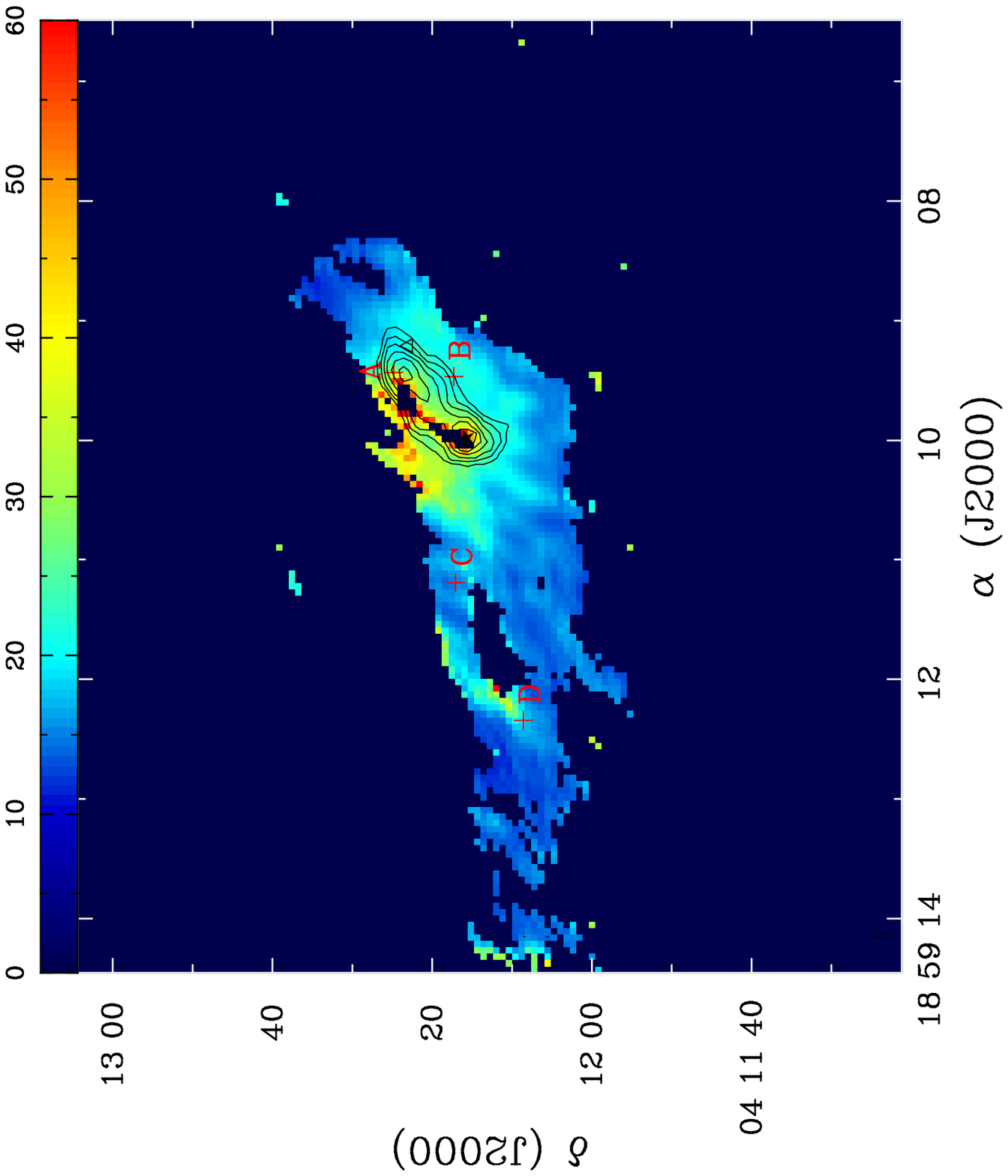}
\caption{The color scale shows the rotation temperatures 
derived from the \nh3 (1,1) and (2,2) lines. The contours represent
the integrated emission of the SiO 2-1 line. The `star' symbol and
`triangle' mark the continuum peaks MM-1 and MM-2, respectively.  The crosses 
mark the position of the hot \nh3 (3,3) features.}
\end{figure*}

\newpage

\begin{figure*}
\label{7}
\includegraphics[width=17cm]{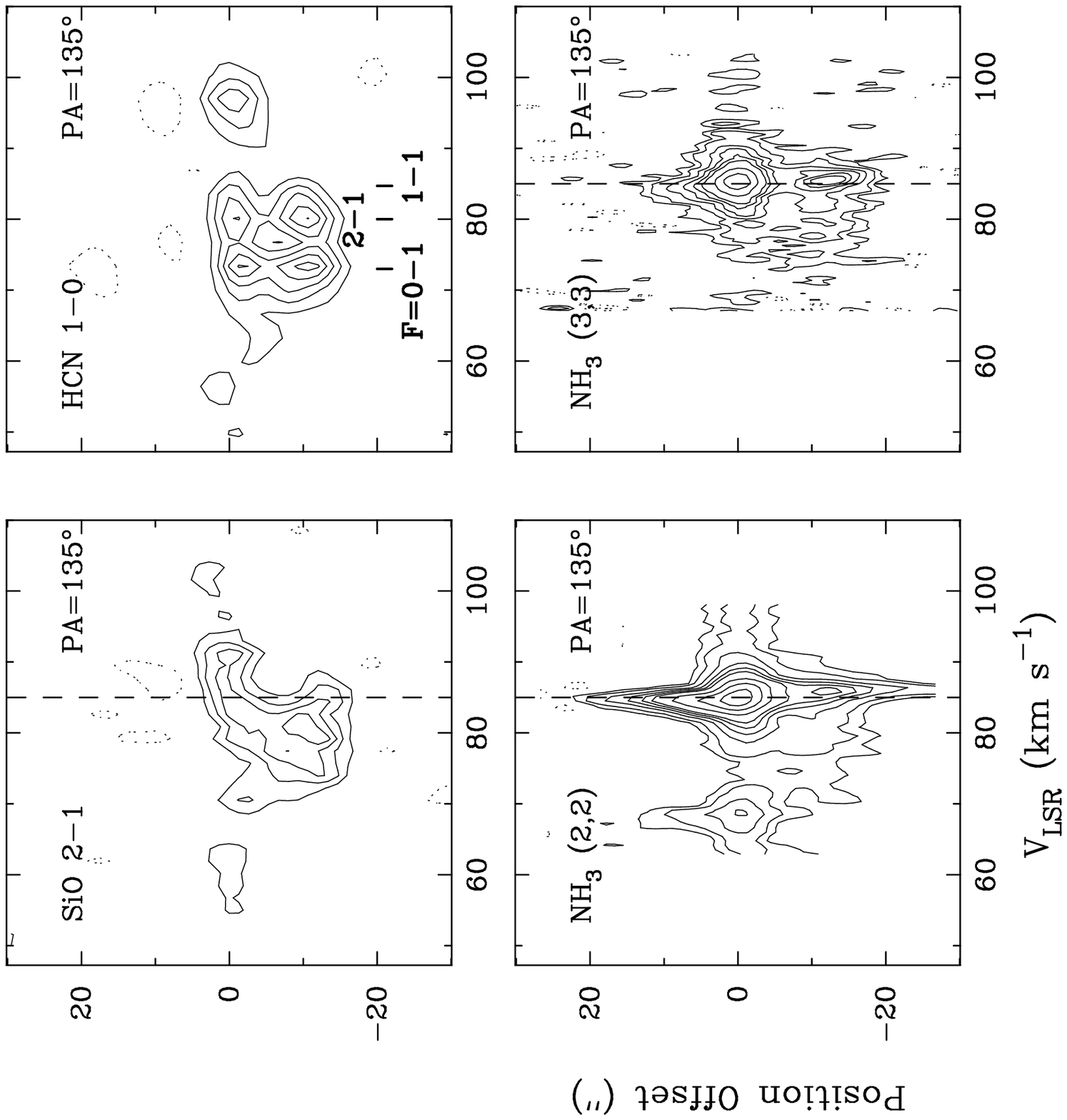}
\caption{The position-velocity plots of the SiO 2-1, HCN 1-0,
\nh3 (J,K)=(1,1), (2,2) and (3,3)
lines. The abcissa is offset from the position of the
dust continuum with position angle (PA) of 135$^\circ$.
The positive offset represents the southeast of the dust continuum peak.
The contours for the SiO and HCN are plotted in steps 
of $\pm$0.05 Jy/beam for the SiO line, and $\pm$0.07
Jy/beam for the HCN line ($\sim 3 \sigma$), respectively. The \nh3 data are 
contoured at 3 $\times$ (1, 2, 4, 6, 8, 10, 15, 20, 25) mJy/beam.
The three HCN hyperfine components are labeled assuming a
cloud systemic velocity of 85.2 \kms-1.
The \nh3 (2,2) emission around 72 \kms-1\ is the satellite hyperfine
component.}
\end{figure*}

\newpage

\begin{figure*}
\label{8}
\includegraphics[width=17cm]{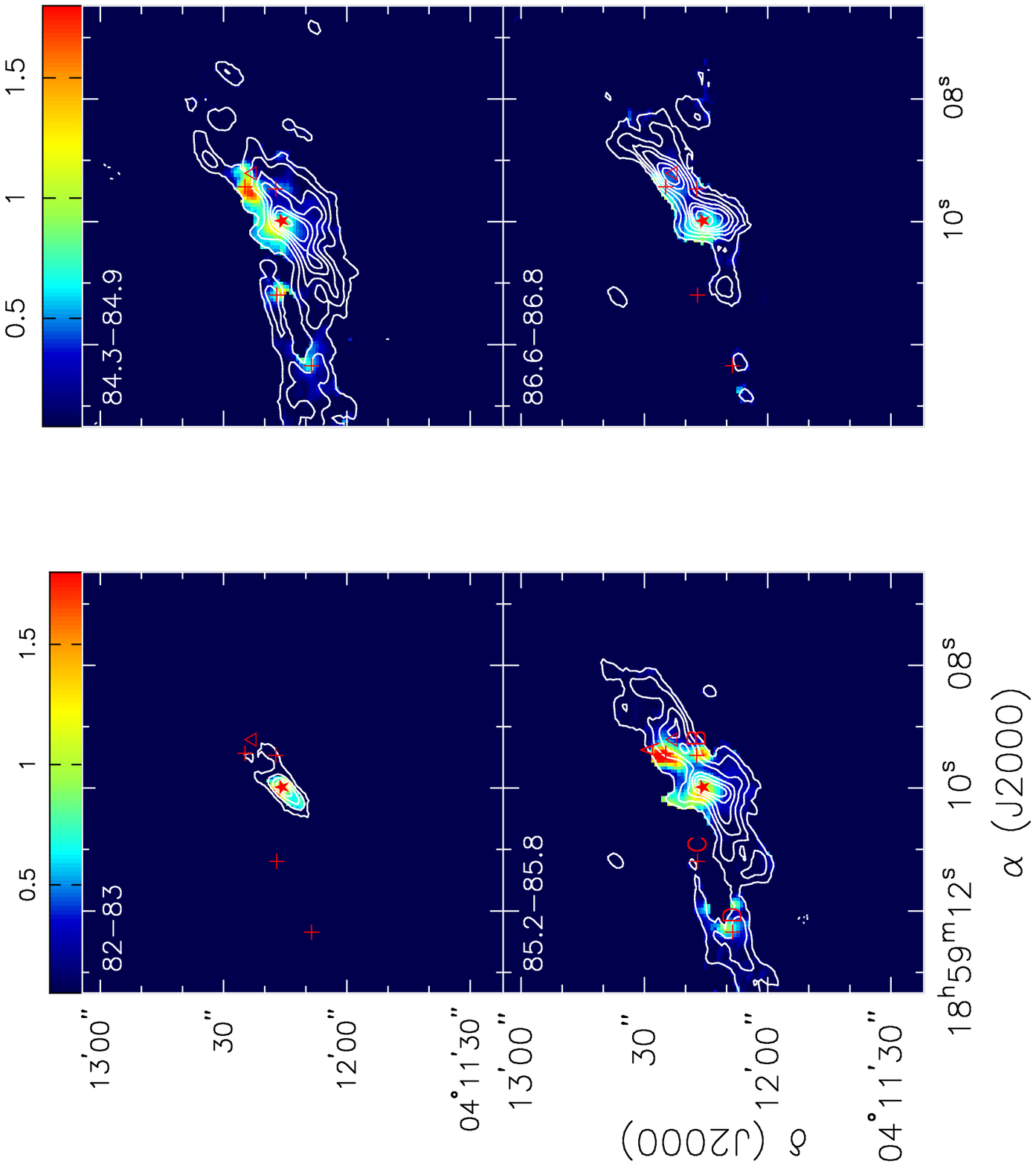}
\caption{The ratio of the integrated \nh3 emission (3,3)/(1,1)
(color) overlaid on the integrated emission of the \nh3 (J,K)=(1,1) 
line (contours). The velocity range of the integration in units
of \kms-1 are given
at the upper left corner of each panel.
The `star' symbol and
`triangle' mark the continuum peaks MM-1 and MM-2, respectively.
The crosses  marks the position of hot \nh3 (3,3) features.}
\end{figure*}

\newpage

\begin{figure*}
\label{9}
\includegraphics[width=17cm]{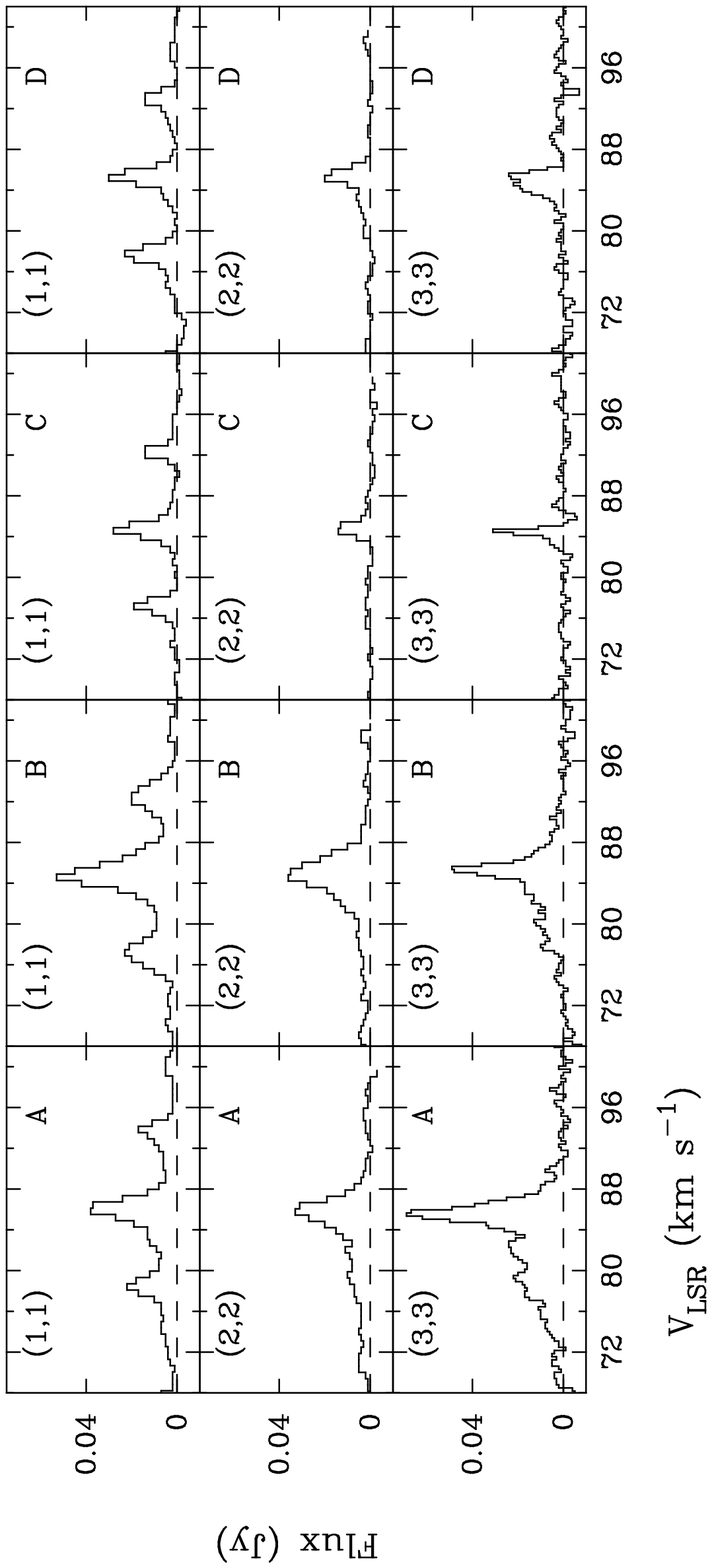}
\caption{Spectra of the \nh3 (J,K)=(1,1), (2,2) and (3,3)
transitions from the compact components A, B, C, and D.}
\end{figure*}


\begin{thebibliography}{}
 
\bibitem{}
Araya, E., Hofner, P., Kurtz, S., Linz, H., Olmi, L., Sewilo, M., 
Watson, C., \& Churchwell, E. 2005, ApJ, 618, 339

\bibitem{}Bachiller, R. 1996, {ARA\&A}, {34}, 111
 
 
\bibitem{BacMarFun93}
Bachiller, R., Mart{\'{\i}}n-Pintado, J., \& Fuente, A. 1993,
{ApJ}, {417}, L45
 
\bibitem[]{}
{Beltr{\'a}n}, M.~T., {Brand}, J., {Cesaroni}, R., 
{Fontani}, F., {Pezzuto}, S., {Testi}, L., \& {Molinari}, S.
2006, A\&A, 447, 221	

\bibitem[]{}
Beuther H., Hunter, T.R., Zhang, Q., Sridharan, T.K., Zhao, J.-H., 
Sollins, P., Ho, P.T.P., Ohashi, N., Su, Y.N., Lim, J., Liu, S.-Y. 
2004, ApJ 616, L23 

\bibitem[]{}
Beuther, H., Schilke, P., Menten, K. M., Motte, F., Sridharan, T. K.,
\& Wyrowski, F. 2002a, ApJ, 566, 945


\bibitem[]{}
Beuther, H., Schilke, P., Sridharan, T. K., Menten, K. M., Walmsley, C. M., \&
Wyrowski, F. 2002b, A\&A, 383, 892

\bibitem[]{}
Beuther, H., Walsh, A., Schilke, P., Sridharan, T. K., Menten, K. M., \&
Wyrowski, F. 2002c, A\&A, 390, 289

\bibitem[]{}
{Beuther}, H., {Schilke}, P., {Gueth}, F., {McCaughrean}, M., 
{Andersen}, M., {Sridharan}, T.~K., \& {Menten}, K.~M. 2002d,
A\&A, 387, 931

\bibitem[]{}
Beuther, H., Zhang, Q., Sridharan, T. K. \&  Chen Y. 2005, ApJ, 628, 800

\bibitem{}
Bronfman, L.,  Nyman, L.-A., \& May, J. 1996, ApJS, 71, 481

\bibitem[Burke \& Hollenbach(1983)]{Burke83} Burke, J.~R., \& 
Hollenbach, D.~J.\ 1983, \apj, 265, 223 

\bibitem{}
Carral, P., Kurtz, S., Rodr{\'\i}guez, L. F., Mart{\'\i}, J.,
Lizano, S., \& Osorio, M. 1999, RMxAA, 35, 97

\bibitem{}
{Cernicharo}, J. and {Castets}, A. and {Duvert}, G., \& {Guilloteau}, S.
1984, A\&A, 139, L13

\bibitem{} 
Cesaroni, R., Felli, M., Jenness, T., Neri, R., Robberto, M.,
  Testi, L., \& Walmsley, C. M. 1999, A\&A, 345, 949
 
 
\bibitem{} 
Cesaroni, R., Felli, M., Testi, L., Walmsley, C. M. \& Olmi, L.
1997, A\&A, 325, 725

\bibitem[]{}
Cesaroni, R., Neri, R., Olmi, L., Testi, L., Walmsley, C.M., \& 
Hofner, P. 2005, A\&A 434, 1039

\bibitem[Churchwell(2002)]{2002ASPC..267....3C} Churchwell, E.\ 2002, Hot 
Star Workshop III: The Earliest Phases of Massive Star Birth, 267, 3

\bibitem[]{}
{Fontani}, F., {Beltr{\'a}n}, M.~T., {Brand}, J., {Cesaroni}, R., 
{Testi}, L., {Molinari}, S., \& {Walmsley}, C.~M. 2005,
A\&A, 432, 921

\bibitem[]{}
Gonzalez-Alfonso, E. \& {Cernicharo}, J. 1993, A\&A,
279, 506
   	
\bibitem[1993]{bonn98} 
Harju, J., Walmsley, C. M. \& Wouterloot, J. G. A.  1993, ApJS,
98, 51

\bibitem{}
Hildebrand, R. H. QJRAS, 1983, 24, 267

\bibitem{HoTow83}
Ho,~P~T.~P., \& Townes, C. H. 1983,
{ARA\&A}, {21}, 239

\bibitem[Hunter et al. (1999)]{1999AJ....118..477H} Hunter, T.~R., Testi, 
L., Zhang, Q., \& Sridharan, T.~K.\ 1999, \aj, 118, 477 

\bibitem{KraJac95}
Kraemer, K. E., \& Jackson, J. M. 1995
{ApJ}, {439}, L9

\bibitem{ManWoo94}
Mangum, J.~G., \& Wootten, A. 1994,
{ApJ}, {433}, L134

\bibitem{MauWilHen86}
Mauersberger, R., Wilson, T.~L., \& Henkel, C. 1986,
{A \& A}, {160}, L13

\bibitem[]{}
McCutcheon, W. H., Dewdney, P. E., Purton, R., \& Sato, T. 1991, AJ, 101, 1435


\bibitem{MadIrvMat86}
Madden S.~C., Irvine W.~M., Matthews H.~E., Brown R.~D., \& Godfrey P.~D. 1986,
{ApJ}, {300}, L79


\bibitem[]{}
Miralles, M. P., Rodr{\'\i}guez, L. F., Scalise, E. 1994, ApJS, 92, 173

\bibitem{}
Molinari, S., Brand, J., Cesaroni, R., \& Palla,  F. 1996, A\&A, 308, 573

\bibitem[]{}
Osterloh, M., Henning, Th., \& Launhardt, R. 1997, ApJS, 110, 71

\bibitem{}
Pineau des For\^ets, G., Flower, D. R., \& Chi\`eze, J.-P. 1997,
in Herbig-Haro Flows and the Birth of Low Mass Stars, eds Bo Reipurth and Claude
Dertout (Dordrecht: Kluwer), p199.

\bibitem[]{}
Sako, S., et al. 2005, Nature, 434, 995

\bibitem[]{}
Shepherd, D. S., \& Churchwell, E. 1996, ApJ, 472, 225


\bibitem[]{}
Shepherd, D.S., Claussen, M.J., \& Kurtz, S. 2001, Science, 292, 1513

\bibitem[]{}
Slysh, V. I., Val'tts, I. E., Kalenskii, S. V., Voronkov, M. A.,
Palagi, F., Tofani, G., \& Catarzi, M. 1999, A\&AS, 134, 115

\bibitem{WalUng83}
Walmsley, C.~M., \& Ungerechts, H. 1983,
{A \& A}, {122}, 164

\bibitem{sco_kwa1976}
{Scoville}, N.~Z. \& {Kwan}, J. 1976, \apj, 206, 718

\bibitem[]{}
{Sridharan}, T.~K., {Beuther}, H., {Schilke}, P., {Menten}, K.~M., \& 
{Wyrowski}, F. 2002, ApJ, 566, 931

\bibitem[]{}
{Walmsley}, C.~M. and {Churchwell}, E. and {Nash}, A., \& {Fitzpatrick}, E.
1982, ApJ, 258, L75

\bibitem[]{}
Williams, S.~J., {Fuller}, G.~A., \& {Sridharan}, T.~K. 2004,
A\&A, 417, 115

\bibitem[]{}
Wilking, B. A., Mundy, L. G., Blackwell, J. H., \& Howe, J. E.
1989, ApJ, 345, 257


\bibitem[]{}
Wu, J. \& Evans, N. J., II 2003, ApJ, 592, L79


\bibitem[]{}
{Zapata}, L.~A., {Rodr{\'{\i}}guez}, L.~F., {Ho}, P.~T.~P., 
{Zhang}, Q., {Qi}, C., \& {Kurtz}, S.~E. 2005, ApJ, 630, L85
	
\bibitem{ZhaHo95}
Zhang, Q., \& Ho,~P.~T.~P. 1995, {ApJ}, {450}, L63

\bibitem[]{}
Zhang, Q., Hunter, T. R., \& Sridharan, T. K. 1998, ApJ, 505, L154
 

\bibitem[]{}
Zhang, Q., Hunter, T. R.,  Sridharan, T. K., \& Cesaroni, R. 1999, ApJ, 527, L117


\bibitem[]{}
Zhang, Q., Hunter, T.R., Sridharan, T.K., \& Ho, P.T.P. 2002, ApJ 566, 982

\bibitem[]{}
Zhang, Q., Hunter, T. R., Brand, J., Sridharan, T.
 K., Molinari, S., Kramer, M. A., \& Cesaroni, R. 2001,
ApJ, 552, L167

\bibitem[]{}
Zhang, Q., Hunter, T. R., Brand, J., Sridharan, T.
 K., Cesaroni, R., Molinari, S., Wang, J., \& Kramer, M. A.  2005,
ApJ, 625, 864

\bibitem{ZiuFriIrv89}
Ziurys, L.~M., Friberg, P., \& Irvine, W.~M. 1989,
{ApJ}, {343}, 201

\end{thebibliography}
\end{document}